\documentstyle[preprint,revtex]{aps}
\begin{document}
\draft
\preprint{IP/BBSR/92-86}
\preprint{}
\begin{title}
\centerline{\bf{GLUON CONDENSATES AT FINITE BARYON}}
\medskip
\centerline{\bf{DENSITIES AND TEMPERATURE}}
 \end{title}
 \author{A. Mishra}
 \begin{instit}
{Physics Department, Utkal University, Bhubaneswar-751004, India}
\end{instit}
\author { H. Mishra and S.P. Misra}
\begin{instit}
{Institute of Physics, Bhubaneswar-751005, India.}
\end{instit}
\begin{abstract}

We derive here the equation
of state for quark matter with a nontrivial vacuum
structure in QCD at finite temperature and
baryon density. Using thermofield dynamics, the parameters of thermal
vacuum and the gluon condensate function are determined through  minimisation
of the thermodynamic potential, along with a self-consistent
determination of the effective gluon and quark masses. The scale parameter
for the gluon condensates is related to the SVZ parameter in the
context of QCD sum rules at zero temperature.
With inclusion of quarks in the thermal vacuum the critical
temperature at which the gluon condensate vanishes
 decreases as compared to that containing only gluons.
At zero temperature, we similarly
obtain the critical baryon density for the same to be about
0.36 fm$^{-3}$.
\end{abstract}
\narrowtext
\section{INTRODUCTION}

Hot dense hadronic matter in relativistic heavy ion collisions
or in early universe is likely to behave as quark gluon plasma.
The study of this in the context of quantum chromodynamics (QCD)
is nontrivial and nonperturbative \cite{nielsen,svz}.
For  gauge fields only we had seen earlier \cite{am91}
that such a vacuum structure with gluon condensates
can emerge dynamically. We examine here the same problem including quarks
at zero or finite temperature, and at finite baryon density, but again
with only gluon condensates \cite{svz,am91}.
We expect that the effect of gluons will dominate vacuum structure
due to the colour factor.

The method used here is variational and nonperturbative,
since only equal time algebra is  taken as input.
The ansatz functions for variation are determined with the minimisation
of energy density at zero temperature, and, of the thermodynamic potential
at finite temperatures and baryon densities. This is a continuation
of the earlier programme \cite{am91} using thermofield dynamics method
(TFD) \cite{tfd} to study the problem at finite temperature and
leads to a simple generalisation of Bogoliubov transformations
as a part of the technology at zero temperature.

We organise the paper as follows. In section II, we consider
quark matter at finite temperature using as stated the method of
thermofield dynamics
\cite{tfd}. In section III, we calculate the
thermodynamic potential for the system of quark matter at finite temperature
and baryon density and then minimise the same
to derive the equation of state. In section IV
we discuss the results.
Similar techniques with explicit ground state construction have also
been applied to study hot nuclear
matter \cite {nmtr} where scalar isoscalar two pion condensates replace
the $\sigma$-field and for chiral symmetry breaking \cite{pot}
with quark condensates.

Nonperturbative nature of QCD has been studied for a long time
in the context of condensates \cite{nielsen,svz,nambu} or scale symmetry
breaking \cite{joglekar} as well as effective field theories
\cite{pot,elis91b} or lattice gauge theories \cite{lat}.
The present approach is complementary to the above.

\section {Quark matter at finite temperature}
We shall very briefly recapitulate the notations of Ref. \cite{am91},
now with quark fields so that finite baryon density effects can be
included.  Let us consider the QCD Lagrangian given as
\begin{equation}
{\cal L} ={\cal L}_{gauge}+{\cal L}_{matter}+{\cal L}_{int},
\end{equation}
\noindent where
\begin{mathletters}
\begin{equation}
{\cal L}_{gauge}=-{1\over 2}G^{a\mu\nu}(\partial_{\mu}{W^{a}}_{\nu}
-\partial_{\nu}{W^{a}}_{\mu}+gf^{abc}{W^{b}}_{\mu}{W^{c}_{\nu})}
+{1\over 4}{G^{a}}_{\mu\nu}{G^{a\mu\nu}},
\end{equation}
\begin{equation}
{\cal L}_{matter}=\bar \psi (i \gamma ^\mu \partial _\mu-m_0)\psi
\end{equation}
\noindent and
\begin{equation}
{\cal L}_{int}=g\bar \psi \gamma ^\mu  \frac {\lambda^a}{2}
W _\mu^a\psi,
\end{equation}
\end{mathletters}
\noindent where ${W^{a}}_{\mu}$ are the SU(3) colour gauge fields.
We shall quantise in Coulomb gauge \cite{schwinger}
and write the
electric field  ${G^{a}}_{0i}$ in terms of
the transverse and longitudinal parts as
\begin{equation}
{G^a}_{0i}=
{^TG^a}_{0i}+{\partial_{i}{f}^{a}},
\end{equation}
where ${f}^{a}$ is to be determined.
We take at time t=0  \cite{am91}
\begin{mathletters}
\begin{equation}
{W^{a}}_{i}(\vec x)={(2\pi)^{-3/2}}\int
{d\vec k\over \sqrt{2\omega(\vec k)}}({a^a}_{i}(\vec k) +
{{a^a}_{i}(-\vec k)}^{\dagger})\exp({i\vec k.\vec x})
\end{equation}
 \noindent and
 \begin{equation}
 {^{T}G^{a}}_{0i}(\vec x)={(2\pi)^{-3/2}} i \int
{d\vec k}{\sqrt{\omega(\vec k)\over 2}}(-{a^a}_{i}(\vec k) +
{{a^a}_{i}(-\vec k)}^{\dagger})\exp({i\vec k.\vec x}),
\end{equation}
\end{mathletters}
where, $\omega(k)$ is arbitrary \cite{schwinger}
and for equal time algebra we have
\begin{equation}
\left[ {a^a}_{i}(\vec k),{{a^b}_{j}(\vec k^{'})}^\dagger\right]=
\delta^{ab}\Delta_{ij}(\vec k)\delta({\vec k}-{\vec k^{'}}),
\end{equation}
with
 \begin{equation}
 \Delta_{ij}(\vec k)={\delta_{ij}}-{k_{i}k_{j}\over k^2}.
\end{equation}
The equal time quantization condition for the fermionic sector
is given as
\begin{equation}
[\psi _{\alpha}^i(\vec x,t),\psi _\beta^j (\vec y,t
)^{\dagger}]_{+}=\delta ^{ij}\delta _{\alpha \beta}\delta(\vec x -\vec y).
\end{equation}
\noindent We now also have the field expansion for fermion field
$\psi$ at time t=0  given as \cite{pot}
\begin{equation}
\psi^{i}(\vec x)=\frac {1}{(2\pi)^{3/2}}\int
\left[U_r(\vec k)c^{i}_{Ir}(\vec k)
+V_s(-\vec k)\tilde c^{i}_{Is}(-\vec k)\right]
e^{i\vec k\cdot \vec x} d\vec k,
\end{equation}
where $U$ and $V$ are given by \cite{spm78}
\begin{equation}
U_r(\vec k)=\left( \begin{array}{c}\cos \frac {\chi (k)}{2}
\\ \vec \sigma \cdot\hat k \sin \frac {\chi (k)}{2}
\end{array}\right)u_{Ir} ;\quad V_s(-\vec k)=\left(
\begin{array}{c}\vec \sigma .\hat k \sin \frac {\chi (k)}{2}
\\ \cos \frac {\chi(k)}{2} \end{array}\right)v_{Is},
\end{equation}
\noindent where the function $\chi(k)$ could be arbitrary.
Here we approximate the same as
$\cos \chi (k)=
{m_Q}/{\epsilon (k)}$ and $\sin \chi(k)={k}/{\epsilon (k)}$,
with $\epsilon (k)=(k^2+m_Q^2)^{1/2}$.
For free fields e.g., $m_Q=m_0$. However, for interacting
fields, we shall determine $m_Q$ in a self consistent manner
as will be discussed later.
The above are consistent with the equal time anticommutation conditions
provided \cite{spm78}
\begin{equation}
[c^{i}_{Ir}(\vec k),c^{j}_{Is}(\vec k')^\dagger]_{+}=
\delta _{rs}\delta^{ij}\delta(\vec k-\vec k')=
[\tilde c_{Ir}^{i}(\vec k),\tilde c_{Is}^{j}(\vec k')
^\dagger]_{+},
\end{equation}
where i and j refer to the colour and flavour indices \cite{pot}.

In Coulomb gauge, the expression for the Hamiltonian
density, ${\cal T}^{00}$ from equation (1) is given as \cite{schwinger}
\begin{eqnarray}
{\cal T}^{00}&=&:{1\over 2}{^{T}{G^a}_{0i}}{^{T}{G^a}_{0i}}+
{1\over 2}{W^a}_{i}(-\vec \bigtriangledown^2){W^a}_{i}+
gf^{abc}W^a_iW^b_j\partial _i W^c_j\nonumber \\ &+&
{{g^2}\over 4}f^{abc}f^{aef}{W^b_i}{W^c_j}{W^e_i}{W^f_j}+
{1\over 2}(\partial_{i}f^{a})(\partial_{i}f^{a})
\nonumber \\ &+&\bar \psi (-i \gamma ^i \partial _i+m_0)\psi
-g\bar \psi \gamma ^\mu  \frac {\lambda^a}{2}
W _\mu^a\psi:,
\label{t00}
\end{eqnarray}
\noindent where : : denotes the normal ordering with respect to
the perturbative vacuum, say $\mid vac>$, defined through
${a^a}_{i}(\vec k)\mid vac>=0$, $c_{Ir}^i(\vec k)\mid vac>=0$
and $\tilde c_{Ir}^i(\vec k)^\dagger\mid vac >=0$.
In order to solve for the operator $f^a$, we first note that
\begin{equation}
f^a=-{W^a}_0-g \; f^{abc}\;{ (\vec \bigtriangledown ^2)}^{-1}
({W^b}_i \; \partial _i {W^c}_0).
\end{equation}
Proceeding as earlier \cite{am91} with a mean field type of approximation
we obtain,

\begin{eqnarray}
\vec \bigtriangledown ^2{W^a}_0 (\vec x )
&&+ g^2 \; f^{abc}f^{cde} \;<vac',\beta\mid  {W^b}_i(\vec x ) \partial _i
(\vec \bigtriangledown ^2)^{-1}({W^d}_j(\vec x ) \mid vac',\beta>
\partial _j{W^e}_0(\vec x ))\nonumber\\ && ={J^a}_0(\vec x ),
\end{eqnarray}
where,
\begin{equation}
J^a_0=gf^{abc}{W^b_i}^{T}{G^c_{0i}}-g\bar \psi \gamma^0 \frac {\lambda ^a}{2}
\psi.
\end{equation}

We note that at zero temperature, $\mid vac';\beta=\infty>=\mid vac'>$
was the nonpertubative ground state as discussed in
Ref. \cite{am91}.
The extra fermionic contributions in eqautions (11) and (14) may be
noted, and thus
the expressions of Ref. \cite{am91}
 will get modified at finite temperatures and densities.
We define  $\mid vac'>$ through a unitary
transformation, in a similar manner to
Gross-Neveu model considered earlier \cite{hm88}, given as
\begin{equation}
\mid vac ^{'}>
=U\mid vac>,
\end{equation}
 \noindent where
 \begin{equation}
 U=\exp({B^\dagger}-B),
\end{equation}
In Ref. \cite{am91}, it was shown that at zero temperature, we may have
\begin{equation}
{B^\dagger}={1\over 2}
\int {f(\vec k){{{a^a}_{i}(\vec k)}^\dagger}
{{{a^a}_{i}(-\vec k)}^{\dagger}}d\vec k},
\end{equation}
\noindent
where $f(\vec k)$ describes gluon condensates.
For the consideration of vacuum destabilisation we should also
take quark condensates \cite{pot}. However, we shall not
do the same here because of the following reason. Our method here shall
consist of minimisation of the thermodynamic potential.
We had earlier determined the temperature dependent mass like
term for the gluon  in a self consistent manner \cite{am91}.
We shall here also need to determine the same
for the quark fields. The numerical computation
with a variation for the vacuum structure of gluon {\em and} quark
condensates becomes forbidding. Besides this, we believe that
the gluons being in
the adjoint representation shall contribute more strongly to QCD interactions,
and therefore dominate the vacuum structure. However, we do include
the effect of quark sector for vacuum structure at finite temperatures
through thermofield dynamics \cite{tfd}.

In the following, we shall consider the effect of temperature as
well as finite baryon density on the behaviour of the gluon
condensates for vacuum structure. We use
the method of thermofield dynamics to consider the above
which is convenient for our purpose while dealing with
operators and expectation values. The thermal average
of an operator is replaced by expectation value in an
extended Hilbert space associated with thermal doubling
\cite {tfd}. For the present case of including temperature
and finite baryon density effects, the thermal vacuum
is given as
\begin{equation}
|vac',\beta>=U_G(\beta)U_Q(\beta)|vac'>
\label{vacbeta}
\end{equation}
 \noindent where $U_G$ and $U_Q$ are unitary operators involving
thermal excitations of gluons and quarks respectively.
For the gluon sector, we have the old expression \cite{am91}
 \begin{equation}
U_G(\beta)
=\exp{({B_G(\beta)}^{\dagger}-B_G(\beta))},
\end{equation}
 with
 \begin{equation}
 {B_G(\beta)}^{\dagger}=
\int \theta(\vec k,\beta){{b^a}_{i}(\vec k)}^{\dagger}
{{{\underline b}^a}_{i}(-\vec k)}^{\dagger}d\vec k.
\end{equation}
In addition, for quark sector we have,
 \begin{equation}
U_Q(\beta)
=\exp{({B_Q(\beta)}^{\dagger}-B_Q(\beta))},
\end{equation}
 with
 \begin{equation}
 {B_Q(\beta)}^{\dagger}=
\int \bigg[\theta_{-}(\vec k,\beta){{c^i}_{Ir}(\vec k)}^{\dagger}
{\underline c}^i_{Ir}(-\vec k)^{\dagger}
+ \theta_{+}(\vec k,\beta){\tilde c}^i_{Ir}(\vec k)^{\dagger}
{\underline {\tilde c}^i}_{Ir}(-\vec k)^{\dagger}
\bigg]\; d\vec k.
\end{equation}

In the above the underlined operators \cite{am91}
correspond to the extra Hilbert space in TFD. Further,
$\theta(\vec k,\beta)$, $\theta_{\pm}(\vec k ,\beta)$ are arbitrary
functions to be determined from the minimisation of the thermodynamic
potential. For example, for free fields these functions are given by
\begin{mathletters}
\begin{equation}
sinh^2\theta(\vec k,\beta) =\frac{1}{\exp(\beta\omega(\vec k,\beta))-1}
\label{distrb}
\end{equation}
and
\begin{equation}
sin^2\theta_{\pm}(\vec k,\beta) =\frac{1}{\exp(\beta
(\epsilon(\vec k,\beta)\pm \mu))+1}
\label{distrf}
\end{equation}
\end{mathletters}
where $\omega(k,\beta)=\sqrt{k^2+m_G^2}$,
$\epsilon(k,\beta)=\sqrt{k^2+m_Q^2}$ and $\mu$ is the quark chemical potential.
However, for interacting fields these will be different and we shall
approximately determine them in a self consistent manner.
We have seen that such a structure as in equation (\ref{vacbeta})
introduces a thermal Bogoliubov transformation
for gluon fields \cite{am91}. We now have, in addition,
the parallel transformation in quark sector given as
\begin{eqnarray}
\left(
\begin{array}{c}
{c^i}_{Ir}(\vec k) \\ {{\underline c^i}_{Ir}(-\vec k)}^{\dagger}
\\ {{\tilde c}^{i}}_{Ir}(-\vec k)
\\ {{\underline{\tilde c}^{i}}_{Ir}(\vec k)}^{\dagger}
\end{array} \right) & = & \left(
\begin{array} {cccc}
\cos\theta_{-}\; &
\;\sin\theta_{-}\; & \;
0 & \; 0 \\
-\sin\theta_{-}\; & \;
\cos\theta_{-}\; & \;
0\; & \;0\\
0 & 0 & \cos\theta_{+} & \sin\theta_{+} \\
0 & 0 & -\sin\theta_+  &\cos \theta_+
\end{array}
\right)
\left(
\begin{array}{c}
{c^i}_{Ir}(\vec k,\beta) \\ {{\underline c^i}_{Ir}(-\vec k,\beta)}^{\dagger}
\\ {{\tilde c}^{i}}_{Ir}(-\vec k,\beta)
\\ {{\underline{\tilde c}^{i}}_{Ir}(\vec k,\beta)}^{\dagger}
\end{array} \right).
\end{eqnarray}

 Our job now is to evaluate the expectation value of
${\cal T}^{00}$ with respect to $\mid vac^{'};\beta>$.
For this purpose, we have the earlier eqautions \cite{am91}
\begin{equation}
<vac';\beta \mid :{W^a}_{i}(\vec x){W^b}_{j}(\vec y):
\mid  vac';\beta >=
{\delta }^{ab}
\times (2  \pi )^{-3}\int d\vec k e^{i\vec k.(\vec x-
\vec y)}\; {F_{+}(\vec  k,\beta )\over \omega (k,\beta )}\;
\Delta _{ij}(\vec k),
\label{ww}
\end{equation}
\begin{eqnarray}
&&{<vac';\beta \mid}: {^{T} G^{a}_{0i}} (\vec x)
{^{T} G^{b}_{0j}} (\vec y):{\mid vac';\beta >} \nonumber\\
&=& \delta ^{ab}\times (2 \pi )^{-3}
\int d{\vec k}e^{i{\vec k}.{(\vec x-\vec y)}}
{\Delta _{ij}(\vec k)\omega (k,\beta )}
F_{-}( k,\beta ).
\label{gg}
\end{eqnarray}
In the above the temperature dependant
 $F_{\pm}(k,\beta )$ are given as
\begin{equation}
F_{\pm}(\vec k,\beta )  = \cosh 2\theta \bigg ({\sinh}^{2}f(k)
\pm{\sinh 2f(k)\over 2}\bigg ) +\sinh ^2 \theta (k,\beta )
\label{fpm}
\end{equation}
where $\sinh ^ 2 \theta (\vec k,\beta)$ is given by equation (\ref{distrb}).
For the quark fields we have the parallel eqautions given as
\begin{mathletters}
\begin{equation}
<\psi^{i}_\alpha(\vec x)^{\dagger}
\psi^{j}_\beta(\vec y)>_{vac',\beta}=
(2\pi)^{-3}\delta^{ij}\int
\Big ( \Lambda _-(\vec k,\beta)\Big )_{\beta\alpha}
e^{-\vec k .(\vec x-\vec y)}d\vec k,
\label{jpj}
\end{equation}
\begin{equation}
<\psi^{i}_\alpha(\vec x)
\psi^{j}_\beta(\vec y)^{\dagger}>_{vac',\beta}=
(2\pi)^{-3}\delta^{ij}\int
\Big ( \Lambda _+(\vec k,\beta)\Big )_{\alpha \beta}
e^{\vec k .(\vec x-\vec y)}d\vec k,
\label{jjp}
\end{equation}
\end{mathletters}
\noindent where
\begin{equation}
\Lambda_{\pm}(\vec  k,\beta)=\mp\frac{1}{2}\Big [\big (\sin ^2 \theta _{-}-
\sin ^2 \theta _{+}\big )+\big (\gamma^0 \cos \chi+
\vec \alpha .\hat k \sin \chi\big)\big (\sin ^2 \theta_{-}
+\sin ^2 \theta_{+}\big )\Big ].
\end{equation}
Using equations (\ref{t00}), (\ref{ww}), (\ref{gg}) and (28),
we then obtain
the expectation value of ${\cal T}^{00}$ with respect to
$\mid vac^{'};\beta>$ as
\begin{eqnarray}
\epsilon_{0}(\beta) &
\equiv & <vac^{'};\beta \mid
:{\cal T}^{00}:\mid vac^{'};\beta> \nonumber \\
& = & C_{F}(\beta)+C_{1}(\beta)+C_{2}(\beta)+{C_{3}(\beta)}^{2}+C_{4}(\beta),
\label{enrgd}
\end{eqnarray}
\noindent where
\begin{mathletters}
\begin{eqnarray}
C_F(\beta)&=&<:\bar \psi (-i\gamma^i\partial _i +
m_0)\psi:>_{vac',\beta}\nonumber\\ &=&
\frac{6}{\pi^2}\int {k^2dk\over \epsilon (k)}
(\sin^2 \theta_{-}(\vec k,\beta)+\sin^2 \theta_{+}(\vec k,\beta))
(m_Qm_{0}+k^2),
\end{eqnarray}
\begin{eqnarray}
C_{1}(\beta) & = & <:{1\over 2}
{^T}{G^a}_{0i}{^T}{G^a}_{0i}:>_{vac^{'};\beta}\nonumber \\
& = & {4\over {\pi^2}}\int \omega(k)k^{2} F_{-}(k,\beta )\;dk,
\end{eqnarray}
\begin{eqnarray}
C_{2}(\beta ) & = & <:{1\over 2}
{W^a}_{i}{(-\vec \bigtriangledown^2)}{W^a}_{i}:>_{vac';\beta }\nonumber
\\ & = & {4\over {\pi^2}}\int {{k^{4}}\over \omega(k)}\;F_{+}(k,\beta )\;dk
\end{eqnarray}
\begin{eqnarray}
  {C_{3}(\beta )}^{2} & = & <:{1\over 4}g^{2}f^{abc}f^{aef}
{W^b}_{i}{W^c}_{j}{W^e}_{i}{W^f}_{j}:>_{vac';\beta }\nonumber \\
& = & \left({{2g}\over {\pi^2}}\int {{k^{2}}\over
{\omega(k,\beta )}}\;F_{+}(k,\beta )\;dk\right)^2 ,
\end{eqnarray}
\noindent and
\begin{eqnarray}
 C_{4}(\beta ) & = & <:{1\over 2}
(\partial_{i}f^{a})(\partial_{i}f^{a}):>_{vac{'};\beta },\nonumber\\
 & = & 4\times (2 \pi )^{-6}\int  d \vec  k {G_1(\vec k,\beta )
+G_2(\vec k,\beta)\over
 {k^2+\phi (k,\beta )}}.
 \end{eqnarray}
 \end{mathletters}
  \noindent  In the above,
\begin{mathletters}
\begin{eqnarray}
G_1(\vec k,\beta ) & = & 3 g^2 \int d  \vec q
 F_{+}({\mid} \vec q\mid,\beta )\;
 F_{-}({\mid} \vec k +\vec q {\mid},\beta  ) \;
{\omega ({\mid  \vec  k +\vec q \mid},\beta )\over
\omega  ({\mid \vec  q\mid},\beta  )}
\nonumber \\
& \times & \Bigl  (1+{{(q^2 +\vec k.\vec q)^2
}\over{q^2(\vec k+\vec q)^2}}\Bigr ),
\end{eqnarray}
\begin{eqnarray}
G_2(\vec k,\beta) & = & -g^2\int d \vec q \Bigg [\Big (
1+\frac{m_Q(\beta)^2}{\epsilon (\vec q,\beta)\epsilon (\vec q-\vec k,\beta)}
+\frac {\vec q . (\vec q-\vec k)}
{\epsilon (\vec q,\beta)\epsilon (\vec q-\vec k,\beta)}\Big )\nonumber \\
& \times & \Big (\sin ^2 \theta_{-}(\vec q,\beta)
\sin ^2 \theta_{-}(\vec q-\vec k,\beta)
+\sin ^2 \theta_{+}(\vec q,\beta )\sin ^2 \theta_{+}(\vec q -\vec k,\beta)\Big
)
\nonumber\\ & - & \Big (
1-\frac{m_Q(\beta)^2}{\epsilon (\vec q,\beta)\epsilon (\vec q-\vec k,\beta)}
-\frac {\vec q . (\vec q-\vec k)}
{\epsilon (\vec q,\beta)\epsilon (\vec q-\vec k,\beta)}\Big )\nonumber \\
& \times &\Big (\sin ^2 \theta_{-}(\vec q,\beta)
\sin ^2 \theta_{+}(\vec q-\vec k,\beta)
+\sin ^2 \theta_{+}(\vec q,\beta )\sin ^2 \theta_{-}(\vec q -\vec k,\beta)\Big
)
\Bigg ]
\end{eqnarray}
\noindent  and
\begin{equation}
\phi  (k,\beta )=  {3g^2\over  {8 \pi  ^2}}
\int {{dk'}\over {\omega  (k',\beta  )}}\;F_{+}(k,\beta )
\biggl (  k^2+{k'}^2-{(k^2-{k'}^2)^2
\over{2kk'}}\log \Big | {{k+k'}\over{k-k'}}
\Big |  \biggr ).
\end{equation}
\end{mathletters}
The expressions above are the same \cite{am91} as earlier, except for
the additional contribution
of $C_F(\beta)$ from the fermionic terms, as well as the fermionic contribution
through $G_2(\vec k,\beta)$ in the expansion for $C_4(\beta)$
arising from the auxiliary fields.

Since for interacting fields, the form of the functions $\omega (\vec k,\beta)$
and $\epsilon (\vec k,\beta)$ are not known, we parametrise them in the free
field form with temperature dependent effective mass parameters for the gluon
and quark
fields given as
\begin{equation}
\omega (\vec k,\beta)=\sqrt {k^2+m_G(\beta)^2};\;\;
\epsilon (\vec k,\beta)=\sqrt {k^2+m_Q(\beta)^2},
\end{equation}
\noindent with $m_G(\beta)$ and $m_Q(\beta)$ are to be calculated
self consistently as below.

We identify the gluon mass $m_G(\beta)$ as earlier \cite {am91,biroo}
through the $self consistency$ $ requirement$ that
\begin{eqnarray}
  m_G(\beta)^2 =
{{2g^2}\over {\pi^2}}\int {{k^{2}}\over
{\omega(k,\beta)}}\;F_{+}(k,\beta)\;dk.\label{mg0}
\end{eqnarray}
This is derived through single contraction  contribution from
the quartic terms \cite{am91}.

We similarly identify the effective quark mass, $m_Q$
from the sum of the single
contractions of the term quartic in the field $\psi$ of ${\cal T}^{00}$
given by equation (\ref{t00}) as well as from the mass term $m_0\bar\psi\psi$.
  Writing the
sum of the single contraction terms of ${{\cal T}^{00}}_{int}$ as
\begin{equation}
{{\bar \psi }^{i}}_{\alpha }(\vec x)M_{\alpha \beta }
({\vec \bigtriangledown }_{x}){\psi ^{i}}_{\beta }(\vec x),
\end{equation}
\noindent we identify the effective quark mass, $m_Q(\beta)$ as
$m_Q(\beta)=m_0+m'_Q(\beta)$, with $m'_Q$ given through the relation
\begin{equation}
{M_{\alpha \beta }}({\vec \bigtriangledown }_{x})
\Big | _{\mid {\vec \bigtriangledown }_{x}
\mid \rightarrow 0}={m'_Q}(\beta) \delta _{\alpha \beta }.
\end{equation}
\noindent Writing $\psi^{i} (\vec x)$ in the momentum space as
\begin{equation}
\psi^{i} (\vec x)=(2\pi)^{-3/2}\int {\tilde \psi }^{i}(\vec k)
exp(i{\vec k}.{\vec x})d{\vec k},
\end{equation}
\noindent we thus have
\begin{eqnarray}
& & {4\over 3}\times{g^{2}\over {(2\pi)^{6}}}
\int \frac {d{\vec k}d{\vec k'}d{\vec {k_1}}}
{(\vec k_1-\vec k')^2+\phi(\vec k_1 -\vec k',\beta)}
{{\bar {\tilde \psi}}^{i,i_f}}_{\alpha}(\vec k)\Big (
\gamma^{0} \Lambda_{+}(\vec {k_1},\beta)
\Big )_{\alpha \beta}{{\tilde \psi}^{i}}_{\beta}(\vec k')
\nonumber\\ &=&(2\pi)^{-3}\int {{\bar {\tilde \psi}}^{i}}_{\alpha}(\vec k)
M_{\alpha \beta}(\vec k'){{\tilde \psi}^{i}}_{\beta}(\vec k')
d{\vec k}d{\vec k'}.
\end{eqnarray}
\noindent Hence, with the identification $M_{\alpha \beta }
(\vec k')\Big | _{\mid {\vec k'}\mid \rightarrow 0}=m'_Q(\beta)\delta_{\alpha
\beta }$, the effective quark mass, $m_Q(\beta)$ as given by
\begin{equation}
m_Q(\beta)=m_0-{{g^2}\over {3\pi^2}}\int dk
\frac{k^2}{k^2+\phi(k,\beta)}\times\frac{m_Q(\beta)}
{\epsilon (k,\beta)}\Big(
\sin ^2 \theta _{-}(\vec k,\beta) +
\sin ^2 \theta _{+}(\vec k,\beta)\Big).
\label{mf0}
\end{equation}
This is the {\em self consistency requirement for the quark mass}, and
is solved
iteratively where the input $m_Q(\beta)$ of the right  hand  side through
$\epsilon(\vec k,\beta)$ becomes equal to the output $m_Q(\beta)$ of the
left hand side.

We shall extremise over the thermodynamic potential
containing $\epsilon_0(\beta)$. For this purpose, as earlier \cite{am91}
we shall take
\begin{equation}
sinh f(\vec k)=Ae^{-Bk^2/2},
\end{equation}
which corresponds to taking a gaussian
distribution for perturbative
gluons in nonperturbative vacuum \cite{am91}.
The energy density,
$\epsilon_{0}(\beta)$ in terms of the dimensionless
quantities $x={\sqrt {B}}k$, $\mu_G={\sqrt B}m_G(\beta)$,
 $\mu_Q={\sqrt B}m_Q(\beta)$
and $y={\beta\over {\sqrt B}}$ then gets parametrised as
\begin{eqnarray}
\epsilon_{0}(A,\beta)
& = & {1\over {B^2}}(I_F(y)+I_{1}(A,y)+I_{2}(A,y)+{I_{3}(A,y)}^{2}+
I_{4}(A,y)) \nonumber \\
& \equiv & {1\over {B^2}}F(A,y),
\end{eqnarray}
\noindent where
\begin{mathletters}
\begin{equation}
I_F(y)=\frac{6}{\pi^2}\int\frac{x^2dx}{\epsilon(x,y)}
(\sin ^2 \theta _{-}(\vec x,y)+\sin ^2 \theta _{+}(\vec x,y))
(\mu_Q(y)\mu_{0}+x^2),
\end{equation}
\noindent  and
\begin{equation}
I_{4}(A,y) =4\times {(2 \pi )^{-6}}\int d{\vec x}
{{G_1(\vec x,y)+G_2(\vec x,y)}\over {x^2+\phi (x,y)}}.
\label{i4}
\end{equation}
\end{mathletters}
\noindent In the above, $G_1(\vec x,y)=G(\vec x,y)$
of \cite{am91}, and
\begin{eqnarray}
G_2(\vec x,y) & = & -g^2\int d \vec x' \Bigg [\Big (
1+\frac{\mu_Q(y)^2}{\epsilon (\vec x',y)\epsilon (\vec x'-\vec x,y)}
+\frac {\vec x' . (\vec x'-\vec x)}
{\epsilon (\vec x',y)\epsilon (\vec x'-\vec x,y)}\Big )\nonumber \\
& \times & \Big (\sin ^2 \theta_{-}(\vec x',y)
\sin ^2 \theta_{-}(\vec x'-\vec x,y)
+\sin ^2 \theta_{+}(\vec x',y )\sin ^2 \theta_{+}(\vec x' -\vec x,y)\Big )
\nonumber \\ & - &
\Big ( 1-\frac{\mu_Q(y)^2}{\epsilon (\vec x',y)\epsilon (\vec x'-\vec x,y)}
-\frac {\vec x' . (\vec x'-\vec x)}
{\epsilon (\vec x',y)\epsilon (\vec x'-\vec x,y)}\Big )\nonumber\\&
 \times & \Big (\sin ^2 \theta_{-}(\vec x',y)
\sin ^2 \theta_{+}(\vec x'-\vec x,y)
+\sin ^2 \theta_{+}(\vec x',y )\sin ^2 \theta_{-}
(\vec x' -\vec x,y)\Big )
\Bigg ].\nonumber\\&
 &
\end{eqnarray}
\noindent  The other expressions $I_1(A,y)$, $I_2(A,y)$, $I_3(A,y)$
and $\phi(x,y)$ are the same as in \cite{am91}. The above
integrals contain $\mu_G(y)$ and $\mu_Q(y)$ which are
determined from the self consistency requirements \cite{am91}
\begin{eqnarray}
\mu_G(y)^2 & =& {{2g^2}\over {\pi^2}}
\int  {{x^2}dx\over \omega(x,y)}
\Biggl [ \biggl (A^{2}e^{-x^2}
+Ae^{-{{x^2}/2}}(1+A^{2}e^{-x^2})
^{1/2}\biggr )
\biggl (1+{2\over \exp({y\omega(x,y)})-1}\biggr ) \nonumber \\ & + &
{1\over \exp({y\omega(x,y)})-1}\Biggr ],
\label{49}
\end{eqnarray}
and
\begin{equation}
\mu_Q(y)=\mu_0-{{g^2}\over {3\pi^2}}\int dx
\frac{x^2}{x^2+\phi(x,y)}\times\frac{\mu_Q(y)}{\epsilon (x,y)}\Big(
\sin ^2 \theta _{-}(x,y)
+\sin ^2 \theta _{+}(x,y) \Big).
\label{muft}
\end{equation}
$\mu_G(y)$ and $\mu_Q(y)$ are solved through an iterative
procedure \cite{am91}.
We shall now calculate the thermodynamic potential and
minimise the same.

\section{Extremisation of thermodynamic potential and results}
At zero temperature, we had considered  extremisation of energy density
to obtain the vacuum structure.
At finite temperatures and baryon densities, the relevant quantity
for extremisation is the thermodynamic potential, which
at temprature $T=1/\beta $ is given as \cite{tfd}

\begin{equation}
{\cal F}(A,\beta )=\epsilon _{0}{(A,\beta)}-{1\over \beta }(S_G+S_F)
-\mu_B \rho_B.
\end{equation}
Here $\mu_B=3 \mu$ is the baryon chemical poential corresponding to
the baryon number density $\rho_B$ given as, with two quark flavours,
\begin{equation}
\rho_B=\frac{1}{3}\times 2\times 3 \times 2\times\frac{1}{(2\pi)^3}
\int \Big (\sin^2\theta_{-}- \sin^2\theta_{+}\Big) d\vec k,
\label{bnb}
\end{equation}
and $\mu$ as in equation (\ref{distrf}) is the quark chemical potential.
Further,
in the above, $\epsilon _{0}(A,\beta)$ is as given in equation (\ref{enrgd}),
and the entropy densities $S_G$ and $S_F$ for the gluon and quark fields
are given as \cite{tfd}
\begin{equation}
S_G=-2\times 8 \times (2\pi)^{-3}
\int d{\vec k}\Big (\sinh ^{2} \theta \; log \bigl(\sinh ^{2} \theta\bigr)
-\cosh ^{2} \theta \; log \bigl(\cosh ^{2} \theta\bigr) \Bigr )
\label{sg}\end{equation}
and
\begin{eqnarray}
S_F&=&-3\times 2\times 2 \times (2\pi)^{-3}
\int d{\vec k}
\Big (\sin ^2 \theta _{-}(\vec k,\beta) log
\big(\sin ^2 \theta _{-}(\vec k,\beta)\big)\nonumber\\
&+&(\cos ^2 \theta _{-}(\vec k,\beta))
log  \big(\cos ^2 \theta _{-}(\vec k,\beta)\big)
+\sin ^2 \theta _{+}(\vec k,\beta)
log  \big(\sin ^2 \theta _{+}(\vec k,\beta)\big)\nonumber\\
&+&(\cos ^2 \theta _{+}(\vec k,\beta)) log
\big(\cos ^2 \theta _{+}(\vec k,\beta)\big)\Big ).
\label{sf}
\end{eqnarray}
Clearly, the factor $2\times 8$ in (\ref{sg}) above comes from the transverse
and colour degrees of freedom for the gluon fields
and the factor $3\times 2\times 2$ in (\ref{sf}) comes from the
colour, flavour and spin degrees of freedom for the quarks as well as
for the antiquarks.
For the minimisation of thermodymic potential we may scale out the
dimensional parameter $1/{B^2}$
and write \cite{am91}
\begin{eqnarray}
{\cal F}(A,\beta )& \equiv &  {1\over {B^2}}{F_1}(A,y)\nonumber
\\ & = &
{1\over {B^2}}\bigg [F(A,y)-{1\over y}({\cal S}_G(A,y)
+{\cal S}_F(A,y))-\mu_B ' \rho_B '\bigg ],
\label{52}
\end{eqnarray}
\noindent where $F(A,y)=B^2 \epsilon _0(A,\beta)$, $\mu_B '=\sqrt{B}\mu_B$,
 $\rho_B '= B^{3/2}\rho_B$
and the entropy densities ${\cal S}_G(A,y)$ and ${\cal S}_F(A,y)$
in dimensionless units are given as
\begin{eqnarray}
{\cal S}_G(A,y)&=&-{8\over {\pi^2}}  \int  x^{2}dx \biggl \{
\bigl ({1\over \exp{(y\omega(x,y))}-1}\bigr )
log \bigl ({1\over \exp{(y\omega(x,y))}-1}\bigr )\nonumber\\ & - &
\bigl (1+{1\over \exp{(y\omega(x,y))}-1}\bigr )
log \bigl (1+{1\over \exp{(y\omega(x,y))}-1}\bigr )\biggr \}
\end{eqnarray}
and
\begin{eqnarray}
{\cal S}_F(A,y)&=&-{6\over {\pi^2}} \int x^{2}dx \biggl \{
\bigl ({1\over \exp{(y(\epsilon(x,y)-\mu)}+1}\bigr )
log \bigl ({1\over \exp{(y(\epsilon(x,y)-\mu)}+1}\bigr )\nonumber\\ & + &
\bigl (1-{1\over \exp{(y(\epsilon(x,y)-\mu)}+1}\bigr )
log \bigl (1-{1\over \exp{(y(\epsilon(x,y)-\mu)}+1}\bigr )
\nonumber\\ & + & \bigl ({1\over \exp{(y(\epsilon(x,y)+\mu)}+1}\bigr )
log \bigl ({1\over \exp{(y(\epsilon(x,y)+\mu)}+1}\bigr )\nonumber\\ & + &
\bigl (1-{1\over \exp{(y(\epsilon(x,y)+\mu)}+1}\bigr )
log \bigl (1-{1\over \exp{(y(\epsilon(x,y)+\mu)}+1}\bigr )
\biggr \}.
\end{eqnarray}
We note that for each $A$, the gluon and quark masses are
determined self consistently through equations (44) and (45)
for the evaluation of the right hand side of equation (50).
We extremise $F_{1}(A,y)$  of equation (\ref{52}) with
respect to the parameter A and obtain the optimum value of A as $A_{min}$ at
a given temperature T and baryon density $\rho_B$.
In fig.1  we plot $A_{min}$ versus T at zero baryon density for $g^2/4\pi=.8$
against the earlier curve which was without
quark excitations in the thermal vacuum.
We find that the critical temperature is about 205 MeV in contrast to
275 MeV of Ref. \cite{am91} where quark excitations were not included.
We have taken here the coupling constant $g^2/4\pi=0.8$ and the
Lagrangian quark mass $m_0$=300 MeV
as a typical value for the constituent quark mass.
Such a result is similar to that of lattice QCD where inclusion of two
flavours of quarks decreases the critical temperature
\cite{unquenched}  to about
150 MeV \cite{gavai} from about 235 MeV of quenched
approximation \cite{quenched}.
For quark mass $m_0$ less than 300 MeV, $T_C$ decreases below 205 MeV,
but for very small quark masses, the numerical calculations tend
to be unreliable.
A$_{min}$ is plotted in fig.2 as a function of
baryon density $\rho _B$ for temperatures 0
and 100 MeV.
We find that A$_{min}$ decreases with increase in
baryon density $\rho _B$,
and vanishes at and above a critical value, $(\rho _B)_{crit}$ of $\rho_B$.
The values of $(\rho_B)_{crit}$
for temperatures 0 and 100 MeV are 0.36 fm$^{-3}$ and 0.88
fm$^{-3}$ respectively.
Clearly in fig. 2  $A_{min}$ decreases with $\rho_B$ and hence in fig. 1
for finite $\rho_B$ the $A_{min}$ versus T curve will always lie
{\em below} the solid curve, yielding a smaller critical temperature.
In fig.3, we plot the effective gluon mass as
a function of $\rho _B $ for the above temperatures.
We see that it starts decreasing with increase in the baryon
density, $\rho_B$,
and for zero temperature, it becomes zero at and above
$(\rho_B)_{crit}$=0.36 fm$^{-3}$ and for T=100 MeV, it decreases and
remains almost a constant above $(\rho_B)_{crit}$=0.88 fm$^{-3}$.
This is because, when the condensate function vanishes, the effective
gluon mass depends only upon the thermal contributions as in
equation (44) and is independent of the baryon density. We then
plot in fig.4, the effective quark mass, $m_Q$ as function of
$\rho _B$, which starts decreasing with increase in $\rho_B$
and there is a discontinuity
at $(\rho_B)_{crit}$ above which the quark mass decreases
slowly. This discontinuity in the mass is associated with
abrupt vanishing of condensate function  with $A_{min}$
approaching zero
at the critical baryon density. Similar behaviour was observed for
the effective nucleon mass by Ellis et al in Ref. \cite {elis91b},
where however the critical baryon density at which the
gluon condensates vanishes was much higher,
being of the order of 2.37 fm$^{-3}$.
This discontinuity might indicate a possible link between vanishing of
gluon condensates and chiral symmetry restoration.

We next estimate the SVZ parameter using the value of A$_{min}$.
This at any temperature is given as
\begin{equation}
\frac{g^2}{4\pi ^2}<:G^a_{\mu\nu}G^{a\mu\nu}:>_{vac'}
=\frac{1}{B^2}\frac{g^2}{\pi^2}\Big [-I_1(A,y)+I_2(A,y)
+I_3(A,y)^2-I_4(A,y)\Big]_{A=A_{min}},
\end{equation}
with $I_1(A,y)$, $I_2(A,y)$, $I_3(A,y)$ and
$I_4(A,y)$ as in equation (41) \cite{am91}.
 In fig.5, we plot the SVZ parameter given above
as a function of $\rho_B$. It starts decreasing
 with increase in $\rho _B$  and for T=0,
becomes minimum at $(\rho_B)_{crit}$ and then increases with
increase in $\rho_B$. It may be noted that although the condensate
function vanishes at $(\rho_B)_{crit}$, the SVZ parameter
is nonzero and increases with density. This is because
the contribution from the fermionic sector to $I_4$ is negative and
increases in magnitude with density. Similar behaviour is also seen for
T=100 MeV except that the magnitude here is smaller as we have
extra positive thermal contributions to $I_1$ and $I_4$.

We also calculate the pressure $P$ given as \cite{fetter}
\begin{equation}
P(\beta )=-{\cal F}(A,\beta )\Big |_{A=A_{min}}.
\end{equation}
In fig.6, we plot the pressure as function of
the baryon number density,
$\rho_B$ for the temperatures 0 and 100 MeV. We see that
the pressure increases with increase in $\rho_B$.
We may compare the present nonperturbative results
with the perturbative estimation
of the pressure \cite{kapusta} given as
\begin{eqnarray}
P_{pert}& = & \frac{1}{2\pi^2}\bigg [ \mu_f k_f(\mu_f^2-2.5m_f^2)
+1.5 m_f^2 ln \left (\frac {\mu_f+k_f}{m_f}\right ) \bigg]\nonumber\\
& - & \frac{\alpha_s}{\pi^3}\bigg [1.5\times\left (\mu_f k_f -m_f^2
ln \big (\frac {\mu_f+k_f}{m_f} \big )\right )^2-k_f^4\bigg ],
\label{pert}
\end{eqnarray}
where $k_f^2=\mu_f^2-m_f^2$ and $\mu_f=\mu_B/3$ is the baryon
chemical potential. In the above we have taken $\alpha_s$=0.8
and considered the case for two flavours of quarks with masses
300 MeV each. This perturbative equation of state
for zero temperature is plotted as the dashed curve in Fig. 6.
As may be seen, the present equation of state is stiffer than
the perturbative equation of state.

\section{discussions}

In the present paper, we have extended the study of nontrivial
ground state structure of QCD with gluon condensates at zero and finite
temperatures \cite{am91} to the case of nonzero baryon densities.
The calculations as earlier are done in Coulomb gauge.
We would have also liked to include the quark condensates
for the vacuum structures in these calculations. It would have
been nice to demonstrate that this effect is not large, as we
believe to be the case. However, simultaneous consideration of gluon
condensates and quark condensates with appropriate self consistency
requirements becomes computationally prohibitive and has not been
attempted here.

The modified gluon
condensate function and the quark distribution
functions are obtained here through minimisation of the thermodynamic
potential with, as stated, a {\em self consistent} determination for the
quark and gluon mass parameters.
At different temperatures, the gluon condensates disappear at
critical values of the baryon density. Inclusion of thermal excitations in
quark sector as here decreases the critical temperature from 275 MeV
to 205 MeV even for zero baryon density, and shall be lower at finite
baryon densities. At zero temperature,
the above critical baryon density is about 0.36/fm$^3$.
The results are similar to that of lattice QCD
\cite{unquenched,gavai,quenched}.

We note that the present approach is based  on QCD Lagrangian of quarks
and  gluons without the introduction of effective fields
\cite{elis91b}, but with a nonperturbative variational approach and
an explicit vacuum structure \cite{am91}. This may
be relevant for quark gluon plasma where baryon structures
are likely to dissolve at high densities or temperatures
\cite{biro,satz,greiner}.
\acknowledgements

The authors are thankful to J. Pasupathy, J.C. Parikh, N. Barik,
Snigdha Mishra, S. N. Nayak and P.K. Panda for many useful
discussions. One of the authors (AM) would like to thank
the Council of Scientific and Industrial Research (C.S.I.R)
for a fellowship. SPM would like to thank Department of
Science and Technology, Government of India for
research grant no SP/S2/K-45/89 for financial assistance.

\def \qcd {G. K. Savvidy, Phys. Lett. 71B, 133 (1977);
S. G. Matinyan and G. K. Savvidy, Nucl. Phys. B134, 539 (1978); N. K. Nielsen
and P. Olesen, Nucl.  Phys. B144, 376 (1978); T. H. Hansson, K. Johnson,
C. Peterson Phys. Rev. D26, 2069 (1982)}

\def \svz {M.A. Shifman, A.I. Vainshtein and V.I. Zakharov,
Nucl. Phys. B147, 385, 448 and 519 (1979);
R.A. Bertlmann, Acta Physica Austriaca 53, 305 (1981)}

\def \nambu{ Y. Nambu, Phys. Rev. Lett. 4, 380 (1960);
Y. Nambu and G. Jona-Lasinio, Phys. Rev. 122, 345 (1961); ibid,
124, 246 (1961);
J.R. Finger and J.E. Mandula, Nucl.Phys.B199, 168 (1982);
A. Amer, A. Le Yaouanc, L. Oliver, O. Pene and
J.C. Raynal, Phys. Rev. Lett. 50, 87 (1983);
ibid, Phys. Rev. D28, 1530 (1983);
 S.L. Adler and A.C. Davis,
Nucl. Phys. B244, 469 (1984); R. Alkofer and P. A. Amundsen,
Nucl. Phys.B306, 305 (1988); A.C. Davis and A.M. Matheson DAMTP 91-34 (1991)}

\def \pot{A. Mishra, H. Mishra and S. P. Misra, Z. Phys. C57, 241 (1993);
A. Mishra and S. P. Misra, Z. Phys. C (To appear)}

\def \lat{K.G. Wilson, Phys. Rev. D10, 2445 (1974); J.B. Kogut, Rev. Mod.
Phys. 51, 659 (1979); ibid 55, 775 (1983); M. Cruetz, Phys. Rev. Lett.
45, 313 (1980); ibid Phys. Rev. D21, 2308 (1980); T. Celik, J. Engels and
H. Satz, Phys. Lett. B129, 323 (1983); H. Satz, in Proceedings of Large Hadron
Collider Workshop, Vol.I, Ed. G. Jarlskog and D. Rein, CERN 90-10, ECFA 90-133
(1990)}

\def \biroo {T. Biro, Ann. Phys. 191, 1 (1989), Phys. Lett B228, 16 (1989);
Phys. Lett. B245, 142 (1990)}
\def \joglekar {J. Collins, A. Duncan and S. Joglekar, Phys. Rev.
D16, 438 (1977); N.K. Nielsen, Nucl. Phys. B120, 212 (1977);
J. Schechter, Phys. Rev. D21, 3393 (1981); A.A. Migdal
and M.A. Shifman, Phys. Lett. 114B, 445 (1982)}

\def \elis {J. Ellis, J.I. Kaputsa and K.A. Olive, Phys. Lett.
B273, 123 (1991); G. E. Brown and Mannque Rho,
Phys. Rev. Lett. 66, 2720 (1991)}

\def \amqcd { A. Mishra, H. Mishra, S.P. Misra and S.N. Nayak,
Pramana (J. of Phys.) 37, 59 (1991); A. Mishra, H. Mishra, S.P. Misra
and S.N. Nayak, Z. Phys. C57, 233 (1993)}

\def \tfd {H. Umezawa, H. Matsumoto and M. Tachiki,{\it
Thermofield Dynamics and Condensed States} (North Holland,
Amsterdam, 1982)}

\def \nmtr {H. Mishra,
 S.P. Misra, P.K. Panda and B.K. Parida, Int. J. Mod. Phys. E1, 405 (1992) }

\def \schwinger{  J. Schwinger, Phys. Rev. 125, 1043 (1962); ibid,
127, 324 (1962); E. S. Abers and B. W. Lee, Phys. Rep. 9C, 1 (1973);
D. Schutte, Phys. Rev. D31, 810 (1985)}

\def \spm { S. P. Misra, Phys. Rev. D18, 1661, 1673 (1978);
A. Le Youanc et al Phys. Rev. Lett. 54, 506
(1985)}

\def \hmgrnv { H. Mishra, S.P. Misra and A. Mishra,
Int. J. Mod. Phys. A3, 2331 (1988);
S.P. Misra, Phys. Rev. D35, 2607 (1987)}

\def \fetter {A. L. Fetter and J. D. Walecka, {\it Quantum Theory of
Many Particle Systems}, McGraw Hill Book Company, 1971}

\def \kapusta {J.I. Kapusta, {\em Finite Temperature
Field Theory }(Cambridge University Press 1989)}

\def \biro {Tamas S Biro, Int. J. Mod. Phys. E1, 39 (1992)}

\def \satz {H. Satz, Nucl. Phys. A498, 495c (1989)}

\def \greiner {P.R. Subramanian, H. Stocker and W. Greiner,
Phys. Lett. B173,  468 (1986)}.
\def \quenched {D. Barkai, K.J.M. Moriarty and C. Rebbi,
Phys. Rev. D30, 1293 (1984); A.D. Kennedy, J. Kuti, S.Meyer and B.J. Pendelton,
Phys. Rev. Lett. 54, 87 (1985)}
\def \unquenched {S. Gottlieb et al, Phys. Rev. D38, 2245 (1988)}
\def \gavai {R.V. Gavai in Proceedings of Workshop on Role of Quark
Matter In Physics and Astrophysics, edited by R.S. Bhalerao and R.V. Gavai,
Bombay 1992}

\newpage
\centerline{\bf Figure Captions}
\bigskip
\noindent {\bf Fig.1:} We plot $A_{min}$
as function of T in MeV . The dashed curve corresponds to the case
of not including thermal excitations in the quark sector. Lowering of
critical temperature with inclusion of quarks in the thermal vacuum
may be noted.
\hfil
\medskip

\noindent {\bf Fig.2:} We plot $A_{min}$
as function of $\rho_B$ in fm$^{-3}$ for temperatures 0 and
100 MeV.
\hfil
\medskip

\noindent {\bf Fig.3:} We plot the gluon mass, $m_{G}$ in MeV
as function of $\rho_B$ in fm$^{-3}$ for temperatures 0 and
100 MeV.
\hfil
\medskip

\noindent {\bf Fig.4:} We plot the fermion mass, $m_{Q}$ in MeV
as function of $\rho_B$ in fm$^{-3}$ for temperatures 0 and
100 MeV. The discontinuity in the same at critical baryon density
may be noted.
\hfil
\medskip

\noindent {\bf Fig.5:} We plot the SVZ parameter in units of
$10^{-2}GeV^4$
as function of $\rho_B$ in fm$^{-3}$ for temperatures 0 and
100 MeV.
\hfil
\medskip

\noindent {\bf Fig.6:} We plot the pressure, P in units of
MeV/fm$^3$
as function of $\rho_B$ in fm$^{-3}$ for temperatures 0 and
100 MeV. The perturbative equation of state as given by
equation (\ref{pert}) for temperature, T=0
is given as the dashed curve of the same figure.
\hfil
\medskip
\end{document}